\documentclass[aps,pra,twocolumn]{revtex4}
\usepackage{graphicx}
\usepackage{dcolumn}
\usepackage{bm}
\usepackage[]{hyperref}
\usepackage{color} 

\begin{document}


\title{Phase-dependent double-$\Lambda$ electromagnetically induced transparency}
\author{Yi-Hsin Chen,$^{1,2}$ Pin-Ju Tsai,$^{1}$ Ite A. Yu,$^{2}$ Ying-Cheng Chen,$^{3}$ and Yong-Fan Chen$^{1,}$}
\email{yfchen@mail.ncku.edu.tw}

\affiliation{$^1$Department of Physics, National Cheng Kung University, Tainan 70101, Taiwan \\
$^2$Department of Physics and Frontier Research Center on Fundamental and Applied Sciences of Matters, National Tsing Hua University, Hsinchu
30013, Taiwan \\
$^3$Institute of Atomic and Molecular Sciences, Academia Sinica, Taipei 10617, Taiwan}

\date{\today}


\begin{abstract}
We theoretically investigate a double-$\Lambda$ electromagnetically induced transparency (EIT) system. The property of the double-$\Lambda$
medium with a closed-loop configuration depends on the relative phase of the applied laser fields. This phase-dependent mechanism differentiates
the double-$\Lambda$ medium from the conventional Kerr-based nonlinear medium, e.g., EIT-based nonlinear medium discussed by Harris and Hau
[Phys. Rev. Lett. 82, 4611 (1999)], which depends only on the intensities of the applied laser fields. Steady-state analytical solutions for the
phase-dependent system are obtained by solving the Maxwell-Bloch equations. In addition, we discuss efficient all-optical phase modulation and
coherent light amplification based on the proposed double-$\Lambda$ EIT scheme.
\end{abstract}




\maketitle

\newcommand{\FigOne}{
    \begin{figure}[t] 
    \includegraphics[width=7cm]{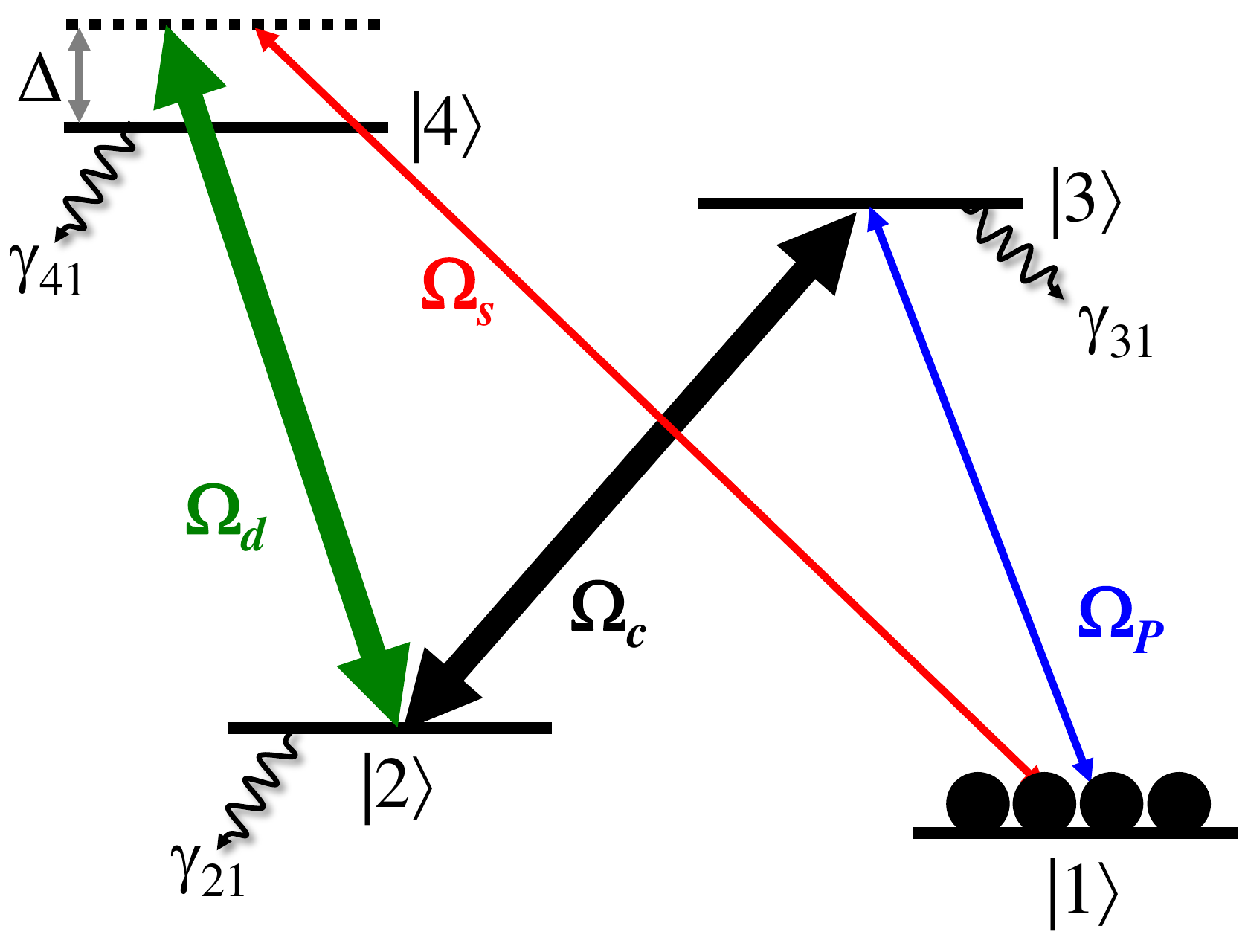}
    \caption{(Color online) Schematic energy level for a double-$\Lambda$ four-level system. States $|1\rangle$ and $|2\rangle$ are two
    metastable ground states, and states $|3\rangle$ and $|4\rangle$ are two excited states. Weak probe ($\Omega_p$) and strong coupling
    ($\Omega_c$) fields form the first EIT system, and weak signal ($\Omega_s$) and strong driving ($\Omega_d$) fields constitute the second EIT system
    with a detuning of $\Delta$. All the atoms are initially prepared in the state $|1\rangle$. The term $\gamma_{31(41)}$ is the total coherence
    decay rate of the excited state $|3\rangle(|4\rangle)$. The term $\gamma_{21}$ is the dephasing rate of the coherence between the ground states $|1\rangle$ and $|2\rangle$.}
    \label{fig:level}
    \end{figure}
}

\newcommand{\FigTwo}{
    \begin{figure}[t] 
    \includegraphics[width=9.0cm]{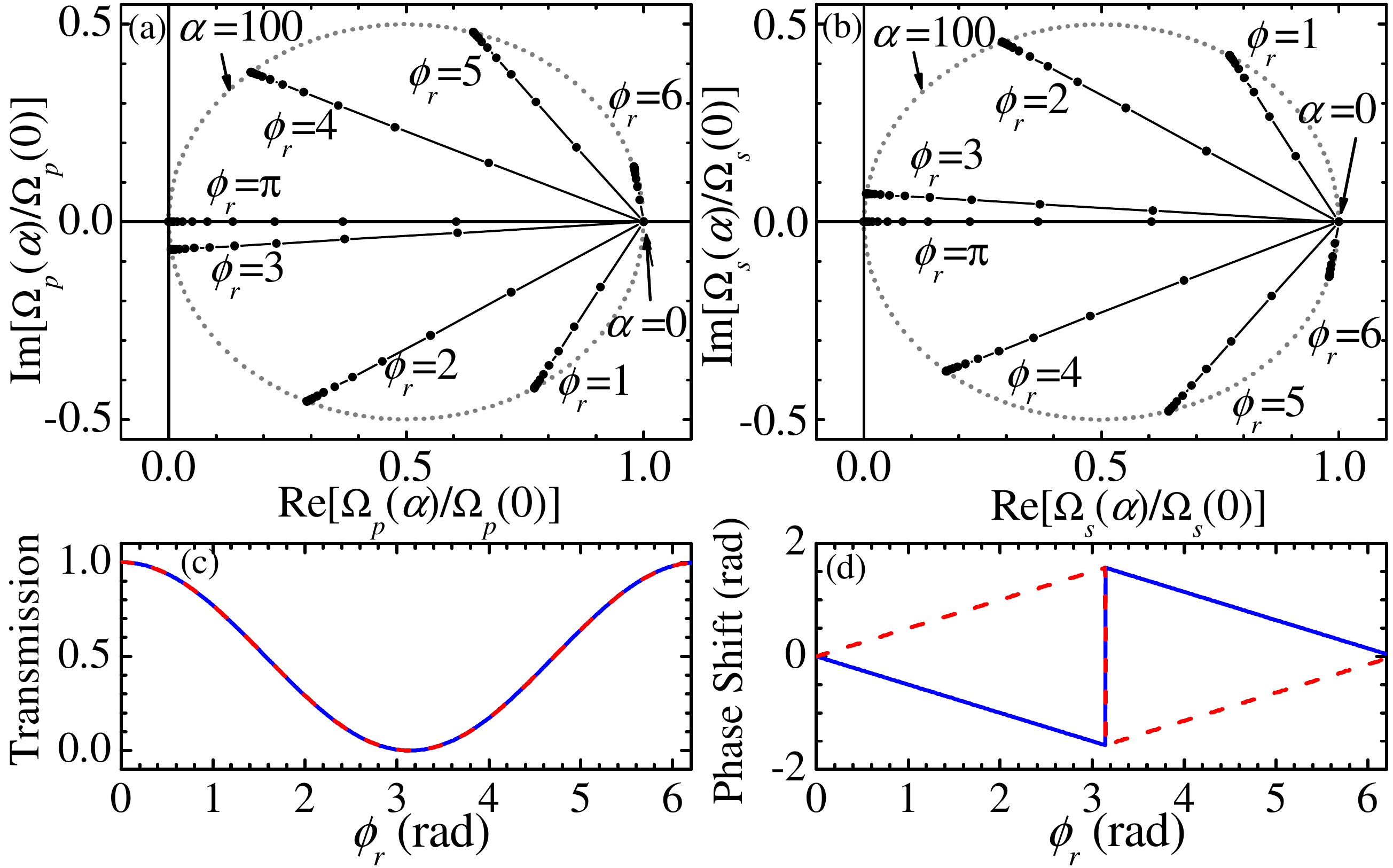}
    \caption{(Color online) Balanced double-$\Lambda$ EIT system ($\Delta=0$, $|\Omega_c|=|\Omega_d|$, and $|\Omega_p(0)|=|\Omega_s(0)|$). (a) and (b)
    show phase diagrams of the probe and signal fields, respectively, plotted according to Eqs.~(\ref{Eq:probesimple}) and (\ref{Eq:signalsimple}). The
    optical depth, $\alpha$, increased from 0 to 100, clearly illustrating the phase evolutions of the light fields. The relative phases, $\phi_r$, are
    set from 1 to 6 as well as $\pi$ in units of rad. The dotted lines show the loops of the light fields at various $\phi_r$ values and $\alpha=100$.
    (c) and (d) are graphs of the corresponding transmission and phase shifts of the probe (blue solid lines) and signal fields (red dashed lines) when
    $\alpha=100$.}
    \label{fig:Fig2}
    \end{figure}
}

\newcommand{\FigThree}{
    \begin{figure}[t] 
    \includegraphics[width=9.0cm]{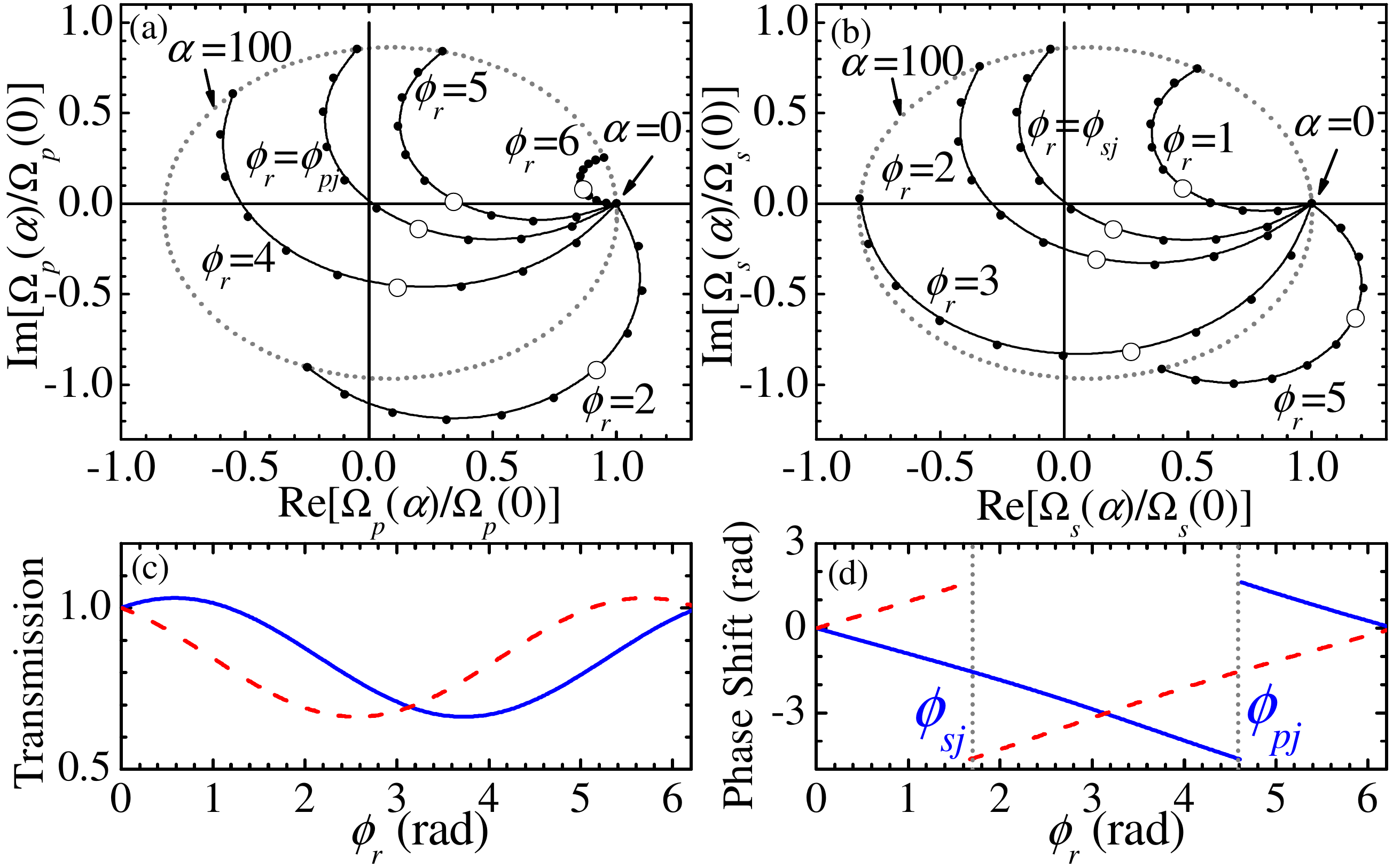}
    \caption{(Color online) Imbalanced double-$\Lambda$ EIT system ($\Delta \neq 0$, $|\Omega_c|=|\Omega_d|$, and $|\Omega_p(0)|=|\Omega_s(0)|$). (a) and
    (b) show phase diagrams of the probe and signal fields, respectively, plotted according to Eqs.~(\ref{Eq:probesimple}) and (\ref{Eq:signalsimple})
    with $\Delta=16.5 \Gamma$. The values of relative phases, $\phi_r$, are set as shown in the plot and $\phi_{pj}=4.62$ and $\phi_{sj}=1.66$ in units of
    rad. (c) and (d) are graphs of the transmission and phase shifts of the probe (blue solid lines) and signal fields (red dashed lines) when
    $\alpha=100$.}
    \label{fig:Fig3}
    \end{figure}
}
\newcommand{\FigFour}{
    \begin{figure}[t] 
    \includegraphics[width=8.0cm]{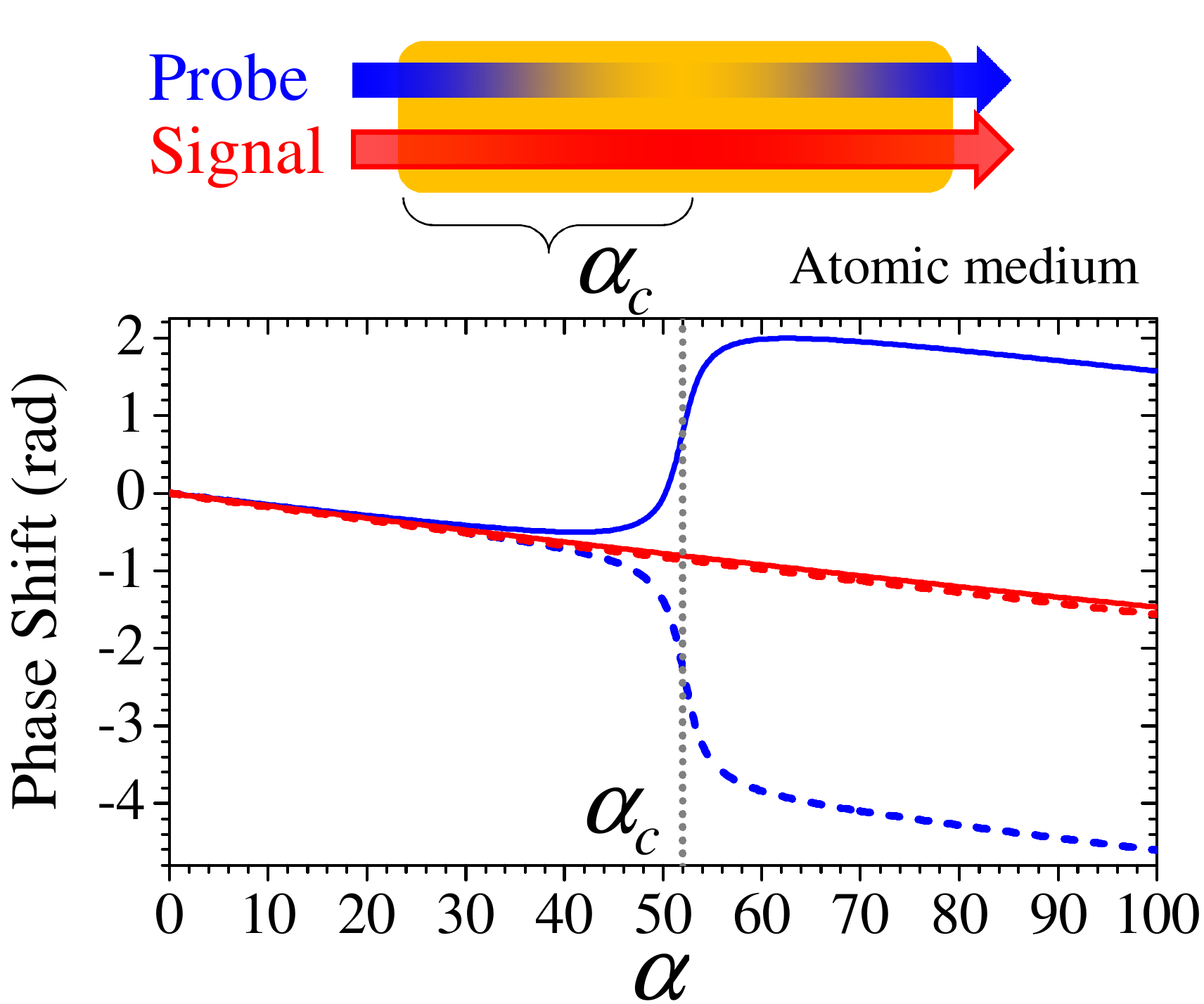}
    \caption{(Color online) The top illustration shows that the probe field exhausts its energy when the light propagates through a medium with a
    critical optical depth of $\alpha_c$. The probe field is then restored when the light field passes through the remainder of medium. The bottom figure
    shows the phase shifts of the probe (blue) and signal (red) fields with the relative phase right above $\phi_{pj}$ (4.67 rad for the solid lines)
    and below $\phi_{pj}$ (4.57 rad for the dashed lines), plotted according to Eqs.~(\ref{Eq:probesimple}) and (\ref{Eq:signalsimple})
    with $\Delta=16.5 \Gamma$.}
    \label{fig:Fig4}
    \end{figure}
}
\newcommand{\FigFive}{
    \begin{figure}[t] 
    \includegraphics[width=8.0cm]{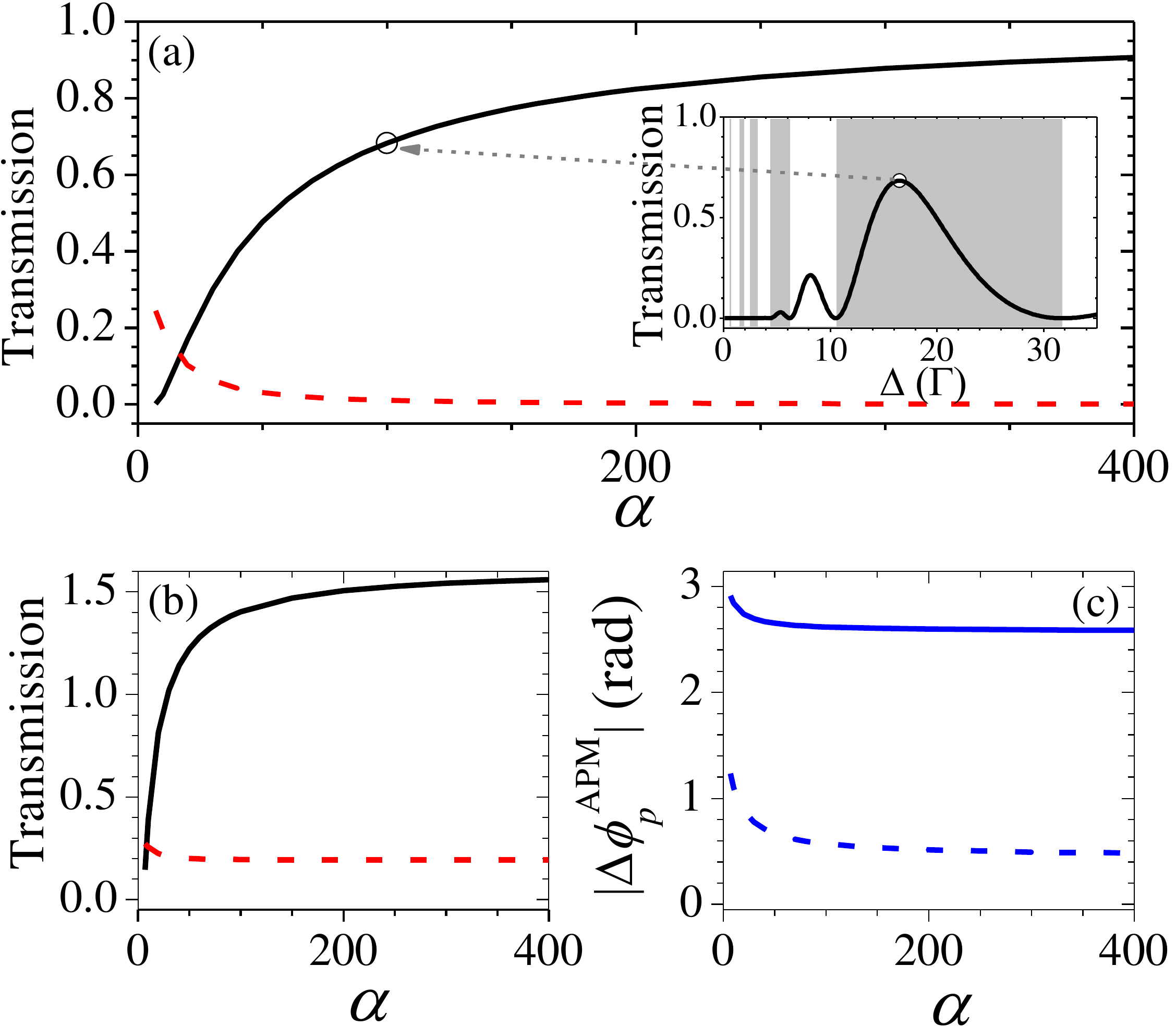}
    \caption{(Color online) (a) Inset: To achieve a $\pi$ phase shift of the transmitted probe field ($|\Delta \phi_p|$) in a double-$\Lambda$ EIT
    system, the probe field transmission is a function of the detuning $\Delta$ with a fixed optical depth ($\alpha=100$). The gray zones in the figure
    show that the terminal point is located on the negative x-axis. A local maximum of transmission is located at approximately $\Delta=16.5 \Gamma$.
    Main plot: Using various optical depths and the corresponding optimized $\Delta$, we obtain the optimized probe transmission, which is a monotonous
    increasing function (black solid line). Without the signal field (red dashed line), the probe transmission is a monotonous decreasing function of
    $\alpha$ with the corresponding $\Delta$. (b) Simulations similar to those shown in (a) except the phase shift of the transmitted probe field
    is set to $\pi/2$. (c) The phase modulation of the probe field with and without the signal field, $|\Delta\phi_p^{\text{APM}}|$, as a function of
    $\alpha$. Blue solid and dashed lines represent the phase modulation in the simulations in (a) and (b), respectively.}
    \label{fig:Fig5}
    \end{figure}
}
\newcommand{\FigSix}{
    \begin{figure}[t] 
    \includegraphics[width=8.0cm]{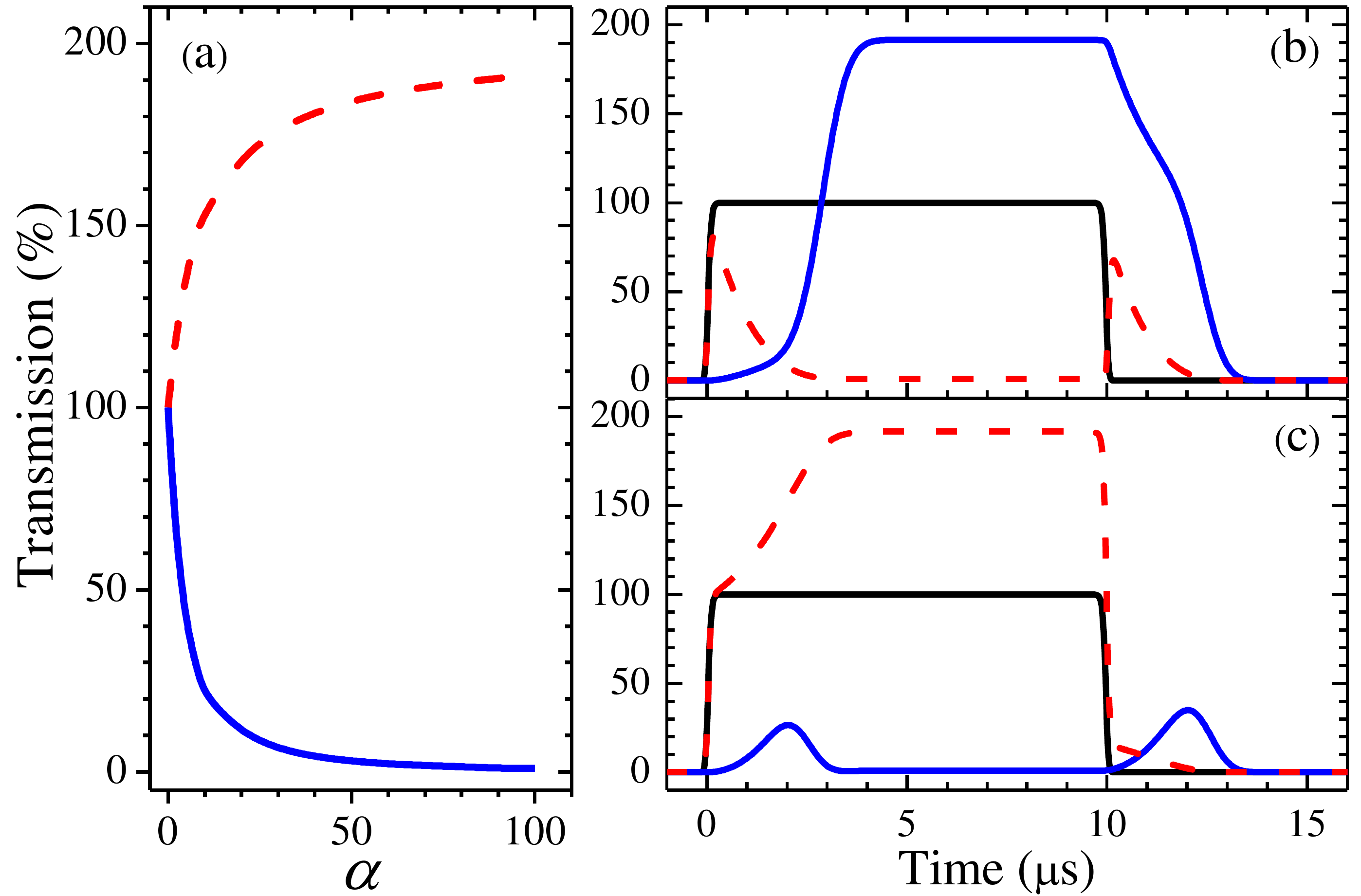}
    \caption{(Color online) (a) Optimal energy amplification for the signal field versus optical depth $\alpha$ with the optimal detuning $\Delta$ and
    relative phase $\phi_r$ determined using numerical simulation of Eqs.~(\ref{Eq:probesimple}) and (\ref{Eq:signalsimple}). Blue solid and red dashed
    lines represent the transmission of the probe and signal fields, respectively. An optical depth of 50 (100) enables achieving an amplification
    efficiency of approximately 84$\%$ (91$\%$), as indicated by the red dashed line. (b) and (c) Two slow-light pulses in a double-$\Lambda$ EIT system
    with relative phases $\phi_r$ of 1.53 rad and 4.76 rad to achieve maximal probe (blue solid lines) and signal transmission (red dashed lines) in (b)
    and (c), respectively. In this simulation, $\alpha=100$, $\Delta=34.2\Gamma$, $\Omega_c=\Omega_d=1\Gamma$, and $\gamma_{21}=0$. Black solid lines
    represent two identical incident probe and signal pulses.}
    \label{fig:Fig6}
    \end{figure}
}


\section{Introduction} \label{Sec:Introduction}

All-optical control of light, including the amplitude and phase of light, is an essential technique in applications for optical and quantum
information science. Electromagnetically induced transparency (EIT) provides an efficient means for manipulating the behavior of light and
coherently controlling photon-photon interaction by using coherent media \cite{Lukin2003,EIT2005}. A light pulse can be slowed and stored in EIT
media \cite{SL99,LsNature,memory13}, prolonging the time of interaction between light and matter and greatly enhancing nonlinearity at low light
levels \cite{Harris99,Chen06}. Recently, efficient all-optical phase modulation (APM) and all-optical switching (AOS) based on EIT have been
proposed and demonstrated \cite{Schmidt96,PsTheory,KangXPM,BrajePS,Chen05,Bajcsy09,Chen10,YFC11}. This EIT-based nonlinearity depends on the
intensities of the applied light fields and the time of interaction between light and matter. Group-velocity-matched double slow-light pulses or
two motionless pulses based on EIT can be applied to enhance the photon nonlinearity, increasing the feasibility of single-photon APM and AOS
\cite{Lukin00,XpmWithSlp,WangDSL,YCCXPM2011,YHCswitching}.


EIT-based four-wave mixing (FWM) can be used to achieve high-efficiency energy conversion between optical fields \cite{Harris99}. A
sum-frequency photon can be efficiently generated with a tunable frequency by using an EIT-based FWM process, forming a two-color slow-light
system \cite{ChenFWM2014}. A two-component or spinor slow light with neutrino-type oscillation has recently been observed in a double-tripod
atom-light coupling scheme \cite{spinorSLTheory,spinorSL}. In this article, we describe a double-$\Lambda$ EIT system with a closed-loop
configuration consisting of four optical fields, as shown in Fig.~\ref{fig:level}. Quantum interference of excitation channels in this system
leads to strong dependence on the relative phase of the four optical fields. The optical properties of the two-color slow light, including the
amplitude and phase, can be controlled by varying the relative phase of the applied laser fields. This phase-dependent mechanism differentiates
the double-$\Lambda$ medium from the conventional Kerr-based nonlinear medium, which depends only on the intensities of the applied laser
fields. By solving the Maxwell-Bloch equations, we obtain steady-state analytical solutions of the double-$\Lambda$ EIT system. Moreover, we
discuss efficient APM and coherent light amplification based on the proposed scheme. Our previous experiment revealed that few-photon
interactions can induce a $\pi$ phase shift using the phase-dependent double-$\Lambda$ EIT system in cold rubidium atoms \cite{YFC13}. In
addition, a phase jump phenomenon was observed in this experiment when the relative phase of the applied laser fields was varied. In this
article, we provide a comprehensive theoretical analysis and a physical explanation and prediction of the phase jump.



\section{Theoretical model} \label{Sec:theory}

We consider a medium consisting of double-$\Lambda$-type four-level atoms with two metastable ground states ($|1\rangle$ and $|2\rangle$) and
two excited states ($|3\rangle$ and $|4\rangle$), as shown in Fig.~\ref{fig:level}. Weak probe (with the Rabi frequency $\Omega_p$) and strong
coupling ($\Omega_c$) fields form the first EIT system, and weak signal ($\Omega_s$) and strong driving ($\Omega_d$) fields constitute the
second EIT system. For an individual EIT system, $\Omega_c$ ($\Omega_d$) manipulates the transmission of $\Omega_p$ ($\Omega_s$) through an
optical dense medium and causes the $\Omega_p$ ($\Omega_s$) to become transparent because of destructive quantum interference. When the
conditions $|\Omega_c| \gg |\Omega_p|$ and $|\Omega_d| \gg |\Omega_s|$ are satisfied, all the atoms remain in the ground state $|1\rangle$ and
the contribution of the probe and signal fields can be treated as a perturbation in the derivation of the following equations. In addition to
the two EIT systems, the double-$\Lambda$ medium can induce two FWM processes: first, $|1\rangle \rightarrow |3\rangle \rightarrow |2\rangle
\rightarrow |4\rangle \rightarrow |1\rangle $, generating the signal field; and second, $|1\rangle \rightarrow |4\rangle \rightarrow |2\rangle
\rightarrow |3\rangle \rightarrow |1\rangle $, generating the probe field. Thus, the energy as well as the phases of the probe and signal fields
are coherently transferred via these two FWM paths. We begin deriving equations from the interaction Hamiltonian between atoms and optical
fields and the equation of the motion of the density matrix operator.  The steady-state analytical solutions for the double-$\Lambda$ EIT system
can be obtained by solving the first-order optical Bloch equations (OBEs) of the atomic density-matrix operator and the Maxwell-Schr\"{o}dinger
equations (MSEs) of the probe and signal pulses as follows:
\begin{eqnarray}
    \frac{d}{dt}\rho_{41} =
    \frac{i}{2}\Omega_{s} + \frac{i}{2}\Omega_{d}\rho_{21} + \left(i\Delta -\frac{\gamma_{41}}{2}\right)\rho_{41},
\label{Eq:OBEp41}\\
    \frac{d}{dt}\rho_{31} = \frac{i}{2}\Omega_{p} + \frac{i}{2}\Omega_{c}\rho_{21} -
    \frac{\gamma_{31}}{2}\rho_{31},
\label{Eq:OBEp31} \\
    \frac{d}{dt}\rho_{21} = \frac{i}{2}\Omega^{\ast}_{c}\rho_{31} +
    \frac{i}{2}\Omega^{\ast}_{d}\rho_{41} - \frac{\gamma_{21}}{2}\rho_{21},
\label{Eq:OBEp21}\\
    \frac{\partial\Omega_{p}}{\partial z} + \frac{1}{c}\frac{\partial\Omega_{p}}{\partial t} = i \frac{\alpha_{p}\gamma_{31}}{2L} \rho_{31},
\label{Eq:MSEprobe}\\
    \frac{\partial\Omega_{s}}{\partial z} + \frac{1}{c}\frac{\partial\Omega_{s}}{\partial t} = i \frac{\alpha_{s}\gamma_{41}}{2L} \rho_{41},
\label{Eq:MSEsignal}
\end{eqnarray}
where $\alpha_{p(s)}$ represents the optical depth of the probe (signal) field transition; $\rho_{ij}$ is the slowly varying amplitude of the
coherence between states $|i\rangle$ and $|j\rangle$; $\gamma_{31(41)} \equiv \Gamma_{3(4)}$ + $\gamma_{3(4)}$ is the total coherence decay rate
of the excited state $|3\rangle$ ($|4\rangle$), where $\Gamma_{3(4)}$ and $\gamma_{3(4)}$ represent the total spontaneous decay rate of the
excited state $|3\rangle$ ($|4\rangle$) and the energy-conserving dephasing rate, respectively \cite{EIT2005}; $\gamma_{21}$ is the dephasing
rate of the coherence between the ground states $|1\rangle$ and $|2\rangle$; $L$ is the optical path length of the medium; and $\Delta$ denotes
the detuning of the signal field transition.

\FigOne

For simplicity, we assume $\alpha_p=\alpha_s\equiv\alpha$, $\gamma_{31}=\gamma_{41}\equiv\Gamma$, and $\gamma_{21}=0$. With the time-derivative
terms being zero, we derive the steady-state solutions by solving the first-order OBEs [Eqs.~(\ref{Eq:OBEp41})--(\ref{Eq:OBEp21})] as follows:

\begin{eqnarray}
&&    \text{$\rho_{21}$}=
    \frac{\text{$\Omega_p$} \text{$\Omega^{\ast}_c$} (\text{$2\Delta+i \Gamma$}) + \text{$\Omega_s$} \text{$\Omega^{\ast}_d$} (\text{$i\Gamma$})}{D},\nonumber \\
&&    \text{$\rho_{31}$}=
    \frac{\text{$\Omega_p$} \text{$|\Omega_d|^2$} - \text{$\Omega_s$} \text{$\Omega_c$} \text{$\Omega^{\ast}_d$}}{D},\nonumber \\
&&    \text{$\rho_{41}$}=
    \frac{\text{$\Omega_s$} \text{$|\Omega_c|^2$} - \text{$\Omega_p$} \text{$\Omega^{\ast}_c$} \text{$\Omega_d$}}{D},
    \label{Eq:rho_ij}
\end{eqnarray}
where $D=-\left[i \Gamma |\text{$\Omega_d$}|^2+(2\Delta +i \Gamma )|\text{$\Omega_c$}|^2\right]$. By substituting Eq.~(\ref{Eq:rho_ij}) into
MSEs [Eqs.~(\ref{Eq:MSEprobe}) and (\ref{Eq:MSEsignal})] with time-derivative components being zero, we obtain the steady-state solutions for
the probe and signal fields as follows:

\begin{eqnarray}
    \text{$\Omega_p(\alpha)$}=\frac{1}{\left|\Omega\right|^2}\left[\left|\text{$\Omega_c $}\right|^2\text{$\Omega_p$}(0)+\text{$\Omega_c$} \text{$\Omega_d^*$} \text{$\Omega_s$}(0)\right] \nonumber \\
    +\frac{1}{\left|\Omega\right|^2}\left[\left|\text{$\Omega_d$}\right|^2\text{$\Omega_p$}(0)-\text{$\Omega_c$} \text{$\Omega_d^*$} \text{$\Omega_s$}(0)\right]e^{-i\frac{\alpha }{2\xi }},
\label{Eq:probe}\\
    \text{$\Omega_s(\alpha)$}=\frac{1}{\left|\Omega\right|^2}\left[\left|\text{$\Omega_d $}\right|^2\text{$\Omega_s$}(0)+\text{$\Omega_d$} \text{$\Omega_c^*$} \text{$\Omega_p$}(0)\right] \nonumber
    \\
    +\frac{1}{\left|\Omega\right|^2}\left[\left|\text{$\Omega_c$}\right|^2\text{$\Omega_s$}(0)-\text{$\Omega_d$} \text{$\Omega_c^*$} \text{$\Omega_p$}(0)\right]e^{-i\frac{\alpha }{2\xi }},
\label{Eq:signal}
\end{eqnarray}
where $|\Omega|^2 = |\Omega_c|^2+|\Omega_d|^2$ and $\xi = i+2\frac{|\text{$\Omega_c$}|^2\Delta}{|\Omega |^2\Gamma}$. The terms $\Omega_p(0)$ and
$\Omega_s(0)$ represent the incident probe and signal fields, respectively. We then consider the phase of each laser field
$\Omega_j=|\Omega_j|e^{i\phi_j}$, where $j=p, s, c ,$ and $d$. The relative phase, $\phi_r$, is defined as ($\phi_p-\phi_c$)--($\phi_s-\phi_d$).
Under the conditions $|\Omega_c|=|\Omega_d|$ and $|\Omega_p(0)|=|\Omega_s(0)|$, we obtain the simple steady-state solutions for the probe and
signal fields as follows:

\begin{eqnarray}
&&    \frac{\text{$\Omega_p(\alpha)$}}{\text{$\Omega_p(0)$}}=
    \frac{1}{2}\left[1+e^{-i\text{$\phi_r$}}+\left(1-e^{-i\text{$\phi_r$}}\right)e^{-i\frac{\alpha }{2\xi }}\right],
\label{Eq:probesimple}\\
&&    \frac{\text{$\Omega_s(\alpha)$}}{\text{$\Omega_s(0)$}}=
    \frac{1}{2}\left[1+e^{i\text{$\phi_r$}}+\left(1-e^{i\text{$\phi_r$}}\right)e^{-i\frac{\alpha }{2\xi }}\right].
\label{Eq:signalsimple}
\end{eqnarray}

The transmission and phase shift of the transmitted probe (signal) field are $|\Omega_{p(s)}(\alpha)/\Omega_{p(s)}(0)|^2$ and
tan$^{-1}$\{Im[$\Omega_{p(s)}(\alpha)$]/Re[$\Omega_{p(s)}(\alpha)$]\}, respectively. According to Eqs.~(\ref{Eq:probesimple}) and
(\ref{Eq:signalsimple}), when $\Delta = 0$ and $\phi_r=\pi$ the double-$\Lambda$ EIT medium becomes opaque and maximally attenuates both the
probe and signal fields. However, when $\phi_r=0$ both the probe and signal fields become completely transparent, as a result of destructive
interference. The phase-dependent double-$\Lambda$ EIT scheme with $\Delta = 0$ can be applied in low-light-level AOS, as previously
described~\cite{ZhuDLambdaPS1}. Theories regarding the influence of the relative phase of the applied laser fields on the transmission of light
fields when propagated through the double-$\Lambda$ medium were discussed in Ref.~\cite{KorsunskyDLambdaT}, and the matched propagation of a
pair of slow light pulses in the double-$\Lambda$ medium was studied in Ref.~\cite{DengDLambdaT}. Here, we discuss efficient APM and coherent
light amplification based on the double-$\Lambda$ EIT scheme.


\section{Results and discussions} \label{Sec:results}

To describe the mechanism and behavior of light pulses propagating in the double-$\Lambda$ EIT medium, we consider the effects of the signal
detuning ($\Delta$), optical depth ($\alpha$), and relative phase ($\phi_r$) in Sections~\ref{Sec:delta0} and \ref{Sec:delta}. The initial
phases of the probe and signal fields ($\phi_p$ and $\phi_s$) are set to 0 in the following calculations. We drew a phase diagram to show the
evolution of the phase shifts and transmission of both the probe and signal fields. The phase diagram reveals that a clear phase jump that
occurs when the relative phase is varied. We discuss the phase jump in Section~\ref{Sec:phasejump}. For practical applications in optical and
quantum control, an efficient APM should satisfy a $\pi$-order phase shift with high light transmission. We vary the parameters for achieving
this main goal and provide a discussion in Section~\ref{Sec:transmission}. Finally, the effect of coherent light amplification and the dynamics
of both the probe and signal pulses propagating in the double-$\Lambda$ EIT medium are presented by numerically simulating
Eqs.~(\ref{Eq:OBEp41})--(\ref{Eq:MSEsignal}) in Section~\ref{Sec:slowlight}. In addition, we show that the steady-state transmission obtained
using the numerical simulations are consistent with those calculated using the analytical solutions [Eqs.~(\ref{Eq:probesimple}) and
(\ref{Eq:signalsimple})].

\FigTwo

\subsection{Balanced double-$\Lambda$ EIT system ($\Delta=0$)} \label{Sec:delta0}

We first discuss a symmetrical double-$\Lambda$ EIT system with a signal detuning of zero ($\Delta=0$). In the case where
$|\Omega_c|=|\Omega_d|$ and $|\Omega_p(0)|=|\Omega_s(0)|$, the two EIT systems are identical and the two FWM paths are balanced. Figures
\ref{fig:Fig2}(a) and \ref{fig:Fig2}(b) show phase diagrams of the probe and signal fields, respectively, plotted according to
Eqs.~(\ref{Eq:probesimple}) and (\ref{Eq:signalsimple}). In the phase diagram, the angle between the x-axis and a line connecting the origin and
data point represents the phase shift, and the square of the distance between the data point and the origin represents the light transmission.
To illustrate the phase evolution of both the probe and signal fields propagating through the double-$\Lambda$ EIT medium, we increase the
optical depth, $\alpha$, from 0 to 100. The relative phase, $\phi_r$, is set from 1 to 6 as well as $\pi$, as shown in Figs. \ref{fig:Fig2}(a)
and \ref{fig:Fig2}(b). The dotted lines show the loops at various $\phi_r$ values and $\alpha=100$. Figures \ref{fig:Fig2}(c) and
\ref{fig:Fig2}(d) show the transmission and phase shifts of the probe (blue solid lines) and signal fields (red dashed lines). When $\Delta=0$,
the two FWM processes in the double-$\Lambda$ EIT system maintain a stable balance; hence, the variations of the probe and signal transmission
according to the relative phase are identical [see Fig. \ref{fig:Fig2}(c)]. Nevertheless, the signs of the variations in the phase shifts of the
probe and signal fields are opposite, as shown in Fig. \ref{fig:Fig2}(d). The phase shifts of the optical fields are continuous variations with
$\phi_r$ and exhibit substantial changes in sign at $\phi_r=\pi$. When $\phi_r < \pi$, the probe (signal) phase shift is negative (positive) and
monotonically decreases (increases) as $\alpha$ increases. By contrast, when $\phi_r > \pi$, the probe (signal) phase shift is positive
(negative) and monotonically increases (decreases) as $\alpha$ increases. When $\phi_r = \pi$, the phase shifts of both the probe and signal
fields are always zero.


\FigThree

\subsection{Imbalanced double-$\Lambda$ EIT system ($\Delta\neq0$)} \label{Sec:delta}

The theoretical analysis revealed that this double-$\Lambda$ EIT scheme is phase dependent. Furthermore, a large phase shift (order $\pi$) of
one weak probe pulse induced by another weak signal pulse can be achieved, then this scheme can be applied in low-light-level APM. However, as
shown in Figs.~\ref{fig:Fig2}(c) and \ref{fig:Fig2}(d), when $\phi_r$ is close to $\pi$, large phase shifts ($\approx \pi/2$) corresponds to low
light transmission ($\approx e^{-\alpha}$), reducing practicality. Therefore, we consider a double-$\Lambda$ EIT system with a non-zero detuning
($\Delta\neq0$), which causes an imbalance between the two FWM processes of the double-$\Lambda$ EIT scheme.

We plotted phase diagrams with $\Delta =16.5 \Gamma$ and $\alpha$ ranging from 0 to 100, as shown in Fig.~\ref{fig:Fig3}. Based on the selected
parameters, high light transmission with a large phase shift can be generated, as discussed in Sec.~\ref{Sec:transmission}. The phase diagrams
show that the phase jump occurs in the probe field when $\phi_r=\phi_{pj}$ and in the signal field when $\phi_r=\phi_{sj}$. Here, we define
$\phi_{pj(sj)}$ as the relative phase when the phase jump occurs in the probe (signal) field. The curve of $\phi_r=\phi_{pj}$ or
$\phi_r=\phi_{sj}$ passing through the origin is a crucial condition for the phase jump [see Figs.~\ref{fig:Fig3}(a) and \ref{fig:Fig3}(b)].
When $\phi_r>\phi_{pj}$ (e.g., $\phi_r=5$ rad), the accumulated phase shift of the transmitted probe field becomes zero at approximately
$\alpha=40$, as indicated by the open circles in Fig.~\ref{fig:Fig3}(a). The probe field then exhibits a constantly increasing positive phase
shift until leaving the medium ($\alpha=100$). By contrast, when $\phi_r<\phi_{pj}$ (e.g., $\phi_r=4$ rad), the probe field constantly increases
negative phase shift throughout propagation. We conclude that the phase shift of the probe (signal) field as a function of $\phi_r$ must become
a phase jump when $\phi_r=\phi_{pj}$ ($\phi_r=\phi_{sj}$), as shown in Fig.~\ref{fig:Fig3}(d).

The double-$\Lambda$ EIT scheme in which $\Delta\neq0$ causes an imbalance between the two FWM paths and leads to that the probe and signal
fields exchange energy mutually. Hence, the energy flow causes the transmission of the probe or the signal field to be greater than unity [see
Fig.~\ref{fig:Fig3}(c)]. In addition, Figs.~\ref{fig:Fig3}(c) and \ref{fig:Fig3}(d) show large phase shifts (order $\pi$) with high light
transmission can be achieved by using the imbalanced double-$\Lambda$ EIT scheme.



\subsection{Phase jump} \label{Sec:phasejump}


The key factor of the phase jump depends on whether the light field disappears during the propagation process (i.e., the curve in the phase
diagram passes through the origin) [see Figs.~\ref{fig:Fig3}(a) and \ref{fig:Fig3}(b)]. In this section, we determine the critical optical depth
($\alpha_c$) and relative phase ($\phi_{pj}$) when the phase jump occurs in the probe field. The curve in the phase diagram terminates at the
origin [i.e., $|\Omega_p(\alpha)/\Omega_p(0)|^2=0$ in Eq.~(\ref{Eq:probesimple})]; hence, we obtain the following equation:

\begin{eqnarray}
    \cot \left(\frac{\phi_{r}}{2}\right)e^{-R}+\tan \left(\frac{\phi_{r}}{2}\right)e^R = -2\sin (I),
\end{eqnarray}
where $R = -\frac{\alpha}{2}\frac{1}{(\Delta /\Gamma )^2+1}$ and $I = \frac{\alpha}{2}\frac{\Delta /\Gamma }{(\Delta /\Gamma )^2+1}$. We define
$\tan (\phi_{r}/2)e^R$ as $\chi$ and simplify the above expression as $\chi ^2+2\sin (I)\chi +1=0$. Hence, we derive

\begin{eqnarray}
    \chi = -\sin (I) \pm i\cos (I).
\end{eqnarray}
Because $\chi$ is a real number, $I=n\pi/2$, where $n$ is an odd integer. We then obtain the analytic solutions of $\alpha_c$ and $\phi_{pj}$ as
follows:

\begin{eqnarray}
&&    \alpha_c=n\pi\frac{(\Delta /\Gamma )^2+1}{\Delta /\Gamma }, \label{Eq:criticalOD}
\end{eqnarray}

\begin{eqnarray}
&&    \phi_{pj}=2 \tan ^{-1}\left[-\sin \left(\frac{n\pi}{2}\right)e^{\frac{n\pi}{2(\Delta /\Gamma )}}\right]. \label{Eq:phipj}
\end{eqnarray}
Similarly, the relative phase of the phase jump for the signal field can be derived from Eq.~(\ref{Eq:signalsimple}) as follows:
\begin{eqnarray}
    \phi_{sj}=2 \tan ^{-1}\left[\sin \left(\frac{n\pi}{2}\right)e^{\frac{n\pi}{2(\Delta /\Gamma )}}\right].
\label{Eq:phisj}
\end{eqnarray}

\FigFour

According to Eqs.~(\ref{Eq:criticalOD})-(\ref{Eq:phisj}) and using $\Delta=16.5\Gamma$ as an example, we obtain $\alpha_c\approx52$,
$\phi_{pj}\approx4.62$ rad, and $\phi_{sj}\approx1.66$ rad for $n=1$; these values are consistent with the numerical results shown in
Fig.~\ref{fig:Fig3}. The phase jump occurs when the light field disappears during the propagation process. As shown in the top plot of
Fig.~\ref{fig:Fig4}, the probe field exhausts its energy when the light propagates through a medium with a critical optical depth ($\alpha_c$),
and the signal field gains the energy. The system is converted into an EIT-based FWM system \cite{ChenFWM2014}. The probe field is then restored
when it passes through the remainder of the medium. We plotted the phase shifts of the probe (blue) and signal (red) fields with the relative
phase slightly above (solid lines) and below $\phi_{pj}$ (dashed lines), as shown in the bottom figure of Fig.~\ref{fig:Fig4}. The figure
clearly illustrates the phase jump near $\phi_{pj}$.

\FigFive

\FigSix

\subsection{All-optical phase modulation} \label{Sec:transmission}

For practicality in optical and quantum control, we discuss a $\pi$-order APM with high light transmission achieved using the double-$\Lambda$
EIT scheme. To achieve a phase shift of $\pi$, the terminal point of the curve in the phase diagram must be located at the negative x-axis
[i.e., the imaginary part of Eq.~(\ref{Eq:probesimple}) is zero and the real part is negative]. We obtain the relative phase for the $\pi$-phase
shift of the probe field as follows:

\begin{eqnarray}
\phi_r^{\pi} =2 \tan^{-1}\left[\frac{\cos(I)-e^{-R}}{\sin(I)}\right]. \label{Eq:phipi}
\end{eqnarray}

In Eq.~(\ref{Eq:probesimple}), $\phi_r$ is substituted by Eq.~(\ref{Eq:phipi}) and we then obtain the light transmission as functions of
$\alpha$ and $\Delta$. We plotted the relationship between the probe field transmission and $\Delta$ with a fixed optical depth ($\alpha=100$)
in the inset of Fig.~\ref{fig:Fig5}(a). The gray zones in the figure show that the terminal point is located on the negative x-axis. The
transmission was maximal at approximately $\Delta=16.5 \Gamma$. Using various optical depths and the corresponding optimized $\Delta$, we obtain
a monotonous increasing function, as indicated by the black solid line in the main plot of Fig.~\ref{fig:Fig5}(a). A $\pi$ phase shift with high
light transmission can be achieved using the double-$\Lambda$ EIT scheme.

To apply the scheme in APM, we compare the light transmission and phase shift with and without the signal field. In an ideal APM technique, a
weak signal pulse can modulate another weak probe pulse by a phase shift of $\pi$ without losing energy. When no $\Omega_s$ is applied in the
proposed scheme, the probe transmission is a monotonous decreasing function of $\alpha$ with the corresponding $\Delta$, as indicated by the red
dashed line in Fig.~\ref{fig:Fig5}(a). Consider the parameters $\alpha=100$, $\Delta=16.5\Gamma$, and the corresponding $\phi_r$ for example.
Although the transmission can be as high as $68\%$ when $\Omega_s$ is present, the transmission becomes only $1\%$ when $\Omega_s$ is absent.
The phase modulation by the signal field, $|\Delta\phi_p^{\text{APM}}|$, is 2.62 rad, as indicated by the black solid line in
Fig.~\ref{fig:Fig5}(c). Hence, achieving a $\pi$-order APM with hight light transmission by using the double-$\Lambda$ EIT scheme remains a
considerable challenge. In addition, we perform a similar simulation except the phase shift of the transmitted probe field, $|\Delta\phi_p|$, is
set to $\pi/2$. In this simulation, the real part of Eq.~(\ref{Eq:probesimple}) is zero and the terminal point in phase diagram is located on
the negative y-axis. As shown in Fig.~\ref{fig:Fig5}(b) and the blue dashed line in Fig.~\ref{fig:Fig5}(c), the probe transmissions are $140\%$
and $19\%$ with and without $\Omega_s$, respectively, and the $|\Delta\phi_p^{\text{APM}}|$ is 0.57 rad when $\alpha=100$.



\subsection{Coherent light amplification} \label{Sec:slowlight}

The phase-dependent double-$\Lambda$ EIT system can coherently convert and amplify the energy of light. The two imbalanced FWM processes cause
two slow-light pulses to exchange energy mutually and the light transmission can be greater than unity. Numerical simulation of optimal energy
amplification for the signal field (i.e., the highest signal transmission) versus optical depth with the optimal detuning and relative phase
revealed that an optical depth of 50 (100) enables attaining an amplification efficiency of approximately 84$\%$ (91$\%$), as indicated by the
red dashed line in Fig.~\ref{fig:Fig6}(a).


We present the dynamics of pulse-shape light propagation by numerically solving MSEs and OBEs. Two identical square pulses are fired into the
medium simultaneously. To clearly observe phase-dependent slow light pulses, we set $\alpha=100$, $\Delta=34.2\Gamma$,
$\Omega_c=\Omega_d=1\Gamma$, and $\gamma_{21}=0$. The relative phases $\phi_r$ of 1.53 rad and 4.76 rad enable achieving maximum probe (blue
solid lines) and signal (red dashed lines) transmission, as shown in Figs.~\ref{fig:Fig6}(b) and \ref{fig:Fig6}(c), respectively. The
steady-state transmission is consistent with the results obtained by calculating Eqs.~(\ref{Eq:probesimple}) and (\ref{Eq:signalsimple}). In
addition, the group velocities differ between the transmitted probe and the signal pulses because $\Delta\neq0$.

\section{CONCLUSION} \label{Sec:conclusion}

We theoretically demonstrated the proposed double-$\Lambda$ EIT system is phase dependent. When the relative phase of the applied light fields
is controlled, two low-light-level optical pulses can exchange their energy and shift phases during propagating through the phase-dependent
medium. The phase diagram reveals that a phase jump occurs when the relative phase is varied. The light pulse exhausts its energy with the
critical optical depth and phase-jump relative phase, which are the key antecedents to the phenomenon. In addition, to apply in APM, a nonlinear
$\pi$-order phase modulation controlled by a light pulse at low light levels can be achieved by using the proposed scheme.



\section*{ACKNOWLEDGEMENTS}

The authors thank Hao-Chung Chen and Yi-Ting Liao for providing helpful discussions. Y.H.C. acknowledges a postdoctoral fellowship from the
National Tsing Hua University. This work was supported by the National Science Council of Taiwan under Grant No. 101-2112-M-006-004-MY3.




\end{document}